\def\<{\langle}
\def\>{\rangle}

\documentstyle[tighten,pra,aps,epsfig,multicol]{revtex}

\title{A Bell's Inequality Test with Entangled Atoms}
\author{Almut Beige,$^{1}$\cite{AB} William J. Munro,$^{2}$ and
Peter L. Knight$^{1}$}
\address{$^{1}$Optics Section, Blackett Laboratory,
Imperial College, London, London SW7 2BZ, England. \\
$^{2}$Center for Laser Science, Department of Physics,
University of Queensland, Brisbane, Australia.}
\date{\today}

\begin{document}

\maketitle 

\begin{abstract}
Previous work on Bell's inequality realised in the laboratory has
used entangled photons. Here we describe how entangled atoms can
violate Bell's inequality, and how these violations can be
measured with a very high detection efficiency. We first discuss a simple scheme based on two-level atoms inside a cavity to prepare the entangled state. 
We then discuss a scheme using four-level atoms, which requires a parameter regime much easier to access experimentally using current technology. As opposed to other schemes, our proposal relies on the presence of finite decay rates and its implementation should therefore be much less demanding. 
\end{abstract}

\vspace*{0.2cm}
\noindent
\begin{multicols}{2}

\section{Introduction}

Bell's inequalities have a central role in tests of quantum mechanics
and relate to the degree of entanglement between subsystems, an
essential resource in quantum information processing. There are a
number of Bell inequalities for two subsystems where each subsystem
contains a qubit of information. For
example, there exist the  original {\it spin} \cite{Bell65}, Clauser
Horne (CH) \cite{Clauser and Horne 74}, Clauser Horne Shimony Holt
(CHSH)\cite{CHSH69} and
information theoretic \cite{Braunstein and Caves 1988} Bell
inequalities, to name but a few. The particular one considered
generally depends on the system under consideration. A scheme may
violate one Bell inequality but not another. Recently an overview of
Bell's inequalities has been given by Peres \cite{Peres}.

A number of experimental tests of Bell's inequality have already been performed \cite{Clauser,Aspect,KwiatI,Gisin,Kwait,Weihs,mandel} using entangled {\em photons}.
In this paper we propose an experimental test of Bell's inequality
on two macroscopically separated {\em atoms}. Each atom possesses a
two-level system with the states $|0 \rangle$ and $|1 \rangle$. 
We describe a scheme which allows us to prepare the atoms in an arbitrary
superposition of a maximally entangled state and a product state
which is of the form
\begin{eqnarray} \label{instate}
|\varphi \> &=& {\alpha \over \sqrt{2}} \, \left( |10\> - |01\>
\right) + \sqrt{1-|\alpha|^2} \, |00\> 
\end{eqnarray}
in a deterministic way.
To do so we make use of a recently proposed idea by Beige {\it
et al.} \cite{ent} of how to manipulate the decoherence-free states
of $N$ atoms inside a cavity. Together with the control over the
prepared state which can be obtained by following a measurement
proposal by Cook \cite{Cook,behe} based on ``electron shelving'' 
this allows us to
investigate, characterise and test Bell's inequality with a very high
precision and detection efficiency. 

The success rate for the preparation of the initial atomic state  (\ref{instate}) will be denoted by $P_0$. If a photon is emitted in the preparation, the scheme fails. If these events are not detected and ignored this leads to a decrease of the observed violation of Bell's inequality. On the other hand, if the scheme succeeds the fidelity of the prepared state is very close to unity. Therefore we estimate, that Bell's inequality is violated as long as the preparation probability exceeds $71\,\%$, if the scheme is intended to prepare the atoms in the maximally entangled state. In this paper we determine $P_0$ and show that it can, in principle, be arbitrarily close to unity. 

Other tests using atoms or ions have been proposed\cite{Cirac,Phoenix,Knight,Fry,Gerry}.
For instance an experiment, based on the proposal by Cirac and Zoller\cite{Cirac}, to entangle two atoms in a cavity has been performed by Hagley {\em et al.} \cite{Hagley}.
Four trapped ions, respectively, have been entangled experimentally in a deterministic fashion by Sackett {\em et al.}\cite{Sackett} following a 
proposal by M{\o}lmer and S{\o}rensen\cite{Molmer}.
But a test of Bell's inequality using atoms has yet to be realised. The main limiting factor in these experiments is {\em dissipation}\cite{Hagley,Sackett}. 
As opposed to this, the scheme proposed here is based on the 
presence of finite decay rates and should therefore be less demanding
experimentally.

The investigation we are examining here is not strictly a
strong \cite{strong} test of quantum mechanics versus local realism
due to the limited spatial separation of the atoms. For a strict test
the scheme would require separating the two atoms by a distance
larger than the speed of light times the measurement time. 
However this atom based experiment closes the detection inefficiency loophole
while the photon experiments close the causality loopholes \cite{Weihs}.
In the scheme we propose, the observable which is expected to violate 
Bell's inequality is measured in {\em each} run of the experiment and
the state of the two atoms can be determined with almost unity efficiency and a very high precision\cite{behe}. Hence this proposed experiment should be seen as complementary to the 
photon experiments.  

The paper is organised as follows. We begin in the next Section with a
description of a simple scheme based on two two-level atoms inside a cavity that can be used to generate the entangled state (\ref{instate}). We describe the single qubit rotation and a way to measure the state of the atoms. The required parameter regime is, however, experimentally demanding. Therefore, in Section III, a scheme is introduced based on two four-level atoms. This system behaves exactly like in the two-level atom case described above and the discussion in Section II is used to obtain the same results. In Section IV we discuss how to test Bell's inequality and for which parameters a violation of the inequality is expected. A final discussion of the results can be found in Section V.

\section{A simple scheme using two-level atoms}

To prepare two two-level atoms in the entangled state (\ref{instate}) they are
placed at fixed positions in a cavity which acts as a resonator for an electromagnetic field. The atoms (or ions) can be stored in the nodes
of a standing light field or in a linear trap.
In the following $|0 \rangle_i$ denotes the ground state and $|1 \rangle_i$ the excited state of atom $i$, respectively, and we assume that the cavity field is in resonance with the atomic transition. We also assume that the coupling constant of each atom with the cavity field is the same and given by $g$, which can be chosen to be real. The cavity should be non-ideal, that is a photon can leak out with a rate $\kappa$ as shown in Fig.~\ref{fig1}. The spontaneous decay rate of each atom equals $\Gamma$.
The distance between the atoms inside the cavity should be much
larger than an optical wavelength. This allows us to address each atom
individually with a laser pulse. The Rabi frequency for atom $i$ will be denoted by $\Omega^{(i)}$ and is in general complex, because we have already chosen $g$ to be real.

\noindent
\begin{minipage}{3.38truein}
\begin{center}
\begin{figure}[h]
\epsfig{file=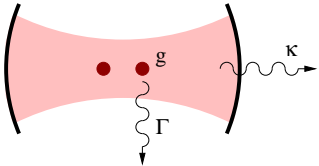,width=4.2cm}\\[0.3cm]
\caption{Experimental setup for the preparation of state
(\ref{instate}). The system consists of two two-level atoms placed
at fixed positions inside a cavity. Each atom couples to the cavity
mode with a constant $g$ and its spontaneous decay rate is given by
$\Gamma$. The rate $\kappa$ corresponds to the leakage of photons
through the cavity mirrors.}\label{fig1}
\end{figure}
\end{center}
\end{minipage}
\vspace*{0.3cm}

To test Bell's inequality the atoms have to be moved out of the cavity. 
This can be done by moving the optical lattice or by applying
an electric field, respectively, if the atoms are inside a linear ion
trap. Another possibility is to let the two atoms fly together
through the cavity field during each run of the experiment.

In the experiment we propose, the probability for spontaneous emission
of a photon or leakage of a photon through the cavity mirrors will be
shown to be small. This immediately suggests that we use the quantum jump approach \cite{HeWi1,HeWi2,HeWi3,HeWi4}. This method
leads to a {\em conditional} Hamiltonian $H_{\rm cond}$ which gives
the time evolu\-tion of the system under the condition of no photon
emission. Due to the non-Hermiticity of $H_{\rm cond}$, the norm of the state vector
\begin{eqnarray} \label{phit}
|\psi^0 (t) \rangle
&=& {\rm e}^{- {\rm i} H_{\rm cond} t/\hbar} |\psi_0 \rangle
\end{eqnarray}
decreases with time and the probability $P_0$ for {\em no} photon
emission up to time $t$ is given by the squared norm
\begin{eqnarray} \label{29}
P_0 (t) &=& \| \, |\psi^0(t) \rangle \, \|^2 ~.
\end{eqnarray}
If no photon is emitted, the state of the system at time $t$ is the
state (\ref{phit}) normalised to unity.

\subsection{The preparation of the entangled state}

To prepare the atoms in state (\ref{instate}) we will take advantage of the fact
that two-level atoms inside a cavity possess {\em trapped} states \cite{alt,trapped1,trapped2,trapped3} which can also be used to obtain an example of a decoherence-free subspace \cite{ent,DFS1,DFS2,DFS3}. If the atoms are in a trapped state they cannot transfer excitation into the resonator field, even if upper levels are populated. Therefore, if the cavity field is empty and spontaneous emission can be neglected no photon can be emitted by the system and the system is in a
{\em decoherence-free} state.

To find the decoherence-free states of the system let us first assume that the two atoms are inside the cavity, but {\em no} laser field is applied. We choose the interaction picture in a way that the atoms and the cavity mode plus environment are considered as the free system. Then the conditional Hamiltonian equals, as in Ref.~\cite{ent,alt},
\begin{eqnarray} \label{Hcond}
H_{\rm cond} &=& {\rm i} \hbar \, g \sum_{i=1}^2 \Big( b \, |1\>_{ii}\<0| 
- {\rm h.c.} \Big) \nonumber \\ 
& & - {\rm i} \hbar \, \Gamma \, \sum_{i=1}^2 |1\>_{ii}\<1|
- {\rm i} \hbar \, \kappa \, b^\dagger b ~,
\end{eqnarray}
where the operator $b$ is the annihilation operator for photons
in the cavity mode.

Decoherence-free states arise if no interaction between the system
and its environment of free radiation fields takes place. If we neglect spontaneous emissions ($\Gamma=0$) this is exactly the case if the cavity mode is empty \cite{ent} and it is $|\psi\>=|0\> \otimes |\varphi\>\equiv |0\varphi\>$. In addition, the systems own time evolution due to the interaction
between the atoms and the cavity mode should not move the state of
the system out of the decoherence-free subspace. Using
Eq.~(\ref{Hcond}) this leads as in Ref.~\cite{ent} to the condition
\begin{eqnarray} \label{DFSphi}
\sum_{i=1}^2 |0\>_{ii}\<1|\varphi\rangle &=& 0 ~,
\end{eqnarray}
where $|\varphi \>$ is the state of the atoms only.
From this condition we find that the decoherence-free states are the
superpositions of the two atomic states $|{\rm g}\> \equiv |00\>$ and 
\begin{eqnarray} \label{a}
|{\rm a} \> &\equiv& (|10\>-|01\>)/\sqrt{2}
\end{eqnarray}
while the cavity mode is empty. 

Once prepared in a decoherence-free state the state of the system does not change in time with respect to the chosen interaction picture. The reason for this is $H_{\rm cond} \, |\psi \rangle =0$ which can be shown by using Eq.~(\ref{Hcond}) and (\ref{DFSphi}). 

To prepare the atoms in state (\ref{instate}) a weak laser pulse can be used.
As in Ref.~\cite{ent} we assume in the following $\Omega^{(1)} \neq \Omega^{(2)}$ and for all non-vanishing Rabi frequencies
\begin{eqnarray} \label{cond}
\Gamma \ll |\Omega^{(i)}| \ll g 
~~~{\rm and}~~ \kappa \sim g ~.
\end{eqnarray}
This corresponds to a strong coupling between the atoms and the cavity mode, while $g$ and $\kappa$ are of the same order of magnitude.
In this parameter regime we can make use of an effect which can easily be understood in terms of the quantum Zeno effect \cite{misra,Itano,zeno}. The reason for this is that the entangled state given in Eq.~(\ref{instate}) corresponds to a decoherence-free state. We assume now that the system is initially in its ground state which is also decoherence-free. If now rapidly repeated measurements are performed on the system of whether the state of the system still belongs to the decoherence-free subspace or not, the laser interaction cannot move the state of the system out of this subspace. Only a time evolution inside the subspace is possible. Hence the laser pulse can introduce entanglement into the system which is not possible in the free atom case. Equivalently we can interpret this inhibition without invoking Zeno effects as a simple consequence of adiabatic elimination using the separation of the frequency scales in Eq.~(\ref{cond}) \cite{ent}. 

Let us define $\Delta T$ as the time in which a photon leaks out through
the cavity mirrors with a probability very close to unity if the
system is initially prepared in a state with no overlap with a
decoherence-free state. On the other hand, a system in a
decoherence-free state will definitely not emit a photon in $\Delta
T$. Therefore the observation of the free radiation field over a time
interval $\Delta T$ can be interpreted as a measurement of
whether the system is decoherence-free or not \cite{messung}. The
outcome of the measurement is indicated by an emission or no
emission of a photon. This interpretation also holds to a very good
approximation in the presence of the laser field because the effect of the laser over a time interval $\Delta T$ can be neglected, which is why condition (\ref{cond}) has been chosen. As it has been shown in Ref.~\cite{messung}, $\Delta T$ is of the order $1/\kappa$ and $\kappa/g^2$ and much smaller than $1/|\Omega^{(\pm)}|$, 
\begin{equation} \label{pm}
\Omega^{(\pm)} \equiv \left(\Omega^{(1)} \pm \Omega^{(2)} \right)/\sqrt{2}~,
\end{equation}
the typical time scale for the laser interaction. Here the system continuously 
interacts with its environment and the system behaves in a very good
approximation like a system under continuous observation
whose time evolution can easily be predicted with the help of the quantum Zeno effect \cite{misra}. 

Using the measurement interpretation one can easily show that the effect of the laser field on the atomic states can be described by the {\em effective} Hamiltonian $H_{\rm eff}$ which equals \cite{ent}
\begin{eqnarray} \label{Heff}
H_{\rm eff} &=& I\!\!P_{\rm DFS} \, H_{\rm cond} \, I\!\!P_{\rm
DFS}
\end{eqnarray}
and where $I\!\!P_{\rm DFS}$ is the projector on the decoherence-free
subspace. To obtain the conditional Hamiltonian of the system in the presence of the laser field the Hamiltonian
\begin{eqnarray} \label{Hlaser}
H_{\rm laser \, I} &=& {\hbar \over 2} \sum_{i=1}^2 \Big( 
\Omega^{(i)} \, |1\>_{ii}\<0| + {\rm h.c.} \Big)
\end{eqnarray}
has to be added to the right hand side of Eq.~(\ref{Hcond}). If we neglect spontaneous emission $(\Gamma=0)$ this leads to
\begin{eqnarray} \label{Heff2}
H_{\rm eff} &=& {\hbar \over 2} \, \Big( \,
\Omega^{(-)} \, |0{\rm a}\>\<0{\rm g}| + {\rm h.c.} \, \Big) ~.
\end{eqnarray}
By solving the corresponding time evolution, one finds that a laser pulse of length $T$ prepares the atoms in the state given in Eq.~(\ref{instate}) with
\begin{eqnarray} \label{alpha}
\alpha &=& - {\rm i} \, {\Omega^{(-)} \over |\Omega^{(-)}|} \,
\sin \left( {|\Omega^{(-)} |T \over 2} \right) ~.
\end{eqnarray}
Varying the length of the laser pulse allows to change arbitrarily
the value of $|\alpha|$ and the amount of entanglement in the system.

The Hamiltonian in Eq.~(\ref{Heff2}) is Hermitian. Therefore the norm
of a vector developing with $H_{\rm eff}$ is not decreasing and in a
first approximation, due to Eq.~(\ref{29}), the emission of photons
can be neglected. To a very good approximation the cavity mode never does become
populated and the success rate of the preparation scheme $P_0$ equals unity.

\noindent
\begin{minipage}{3.38truein}
\begin{center}
\begin{figure}[h]
\epsfig{file=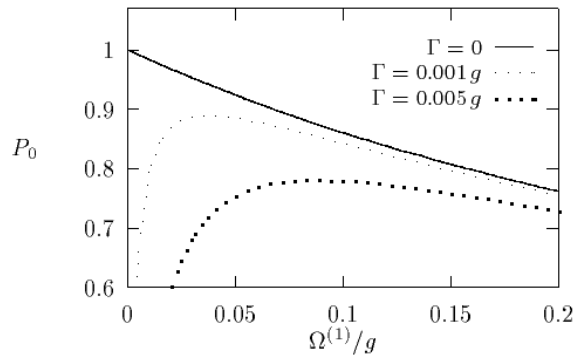,width=8cm}\\[0.3cm]
\caption{The probability for no photon emission during the
preparation of the maximally entangled state for different Rabi
frequencies $\Omega^{(1)}$ and $\Omega^{(2)}=-\Omega^{(1)}$, different
spontaneous decay rates $\Gamma$ and $\kappa=g$.}\label{fig2}
\end{figure}
\end{center}
\end{minipage}
\vspace*{0.5cm}

Fig.~\ref{fig2} shows the probability for no photon emission during the state preparation resulting from a numerical solution of the conditional time
evolution of the system using Eq.~(\ref{29}), (\ref{Hcond}) and (\ref{Hlaser}). This agrees very well with the approximative results given above. As an example, we assumed 
\begin{equation} \label{T}
T =\pi/ |\Omega^{(-)}| ~,
\end{equation} 
which leads, due to Eq.~(\ref{alpha}), to
the preparation of the maximally entangled state of both atoms. In
addition we assumed $\Omega^{(2)}=-\Omega^{(1)}$ \cite{rab}. As expected, for
$\Gamma=0$ the success rate of the preparation scheme can at least in
principle be arbitrarily close to 1. For $\Gamma \neq 0$ the
probability $P_0$ reaches a maximum value for a certain Rabi frequency $\Omega^{(1)}$, but is always smaller than 1. To improve the experiment one can surround the cavity by detectors and repeat it were a decay photon to be registered.

We also determined the state of the atoms at the end of the laser
pulse numerically. The fidelity of the prepared state $F$ in case of
no photon emission is given by the overlap of the state of Eq.~(\ref{phit})
after normalisation with the state given in Eq.~(\ref{instate}). For
the parameters chosen in Fig.~\ref{fig2}, $F$ is found to be always
higher than 95$\,\%$.

\subsection{Realisation of a single qubit rotation}

In this subsection we describe how the single qubit rotation
on atom $i$, defined by the operator $U^{(i)}_{\rm rot}$,
\begin{eqnarray} \label{urot}
U^{(i)}_{\rm rot} (\xi, \phi) & \equiv& \cos \xi
-{\rm i} \, \sin \xi \Big( \, {\rm e}^{{\rm i} \phi} \, |0\>_{ii}\< 1|
+ {\rm h.c.} \, \Big) ~,
\end{eqnarray}
can be realised, where $\xi$ and $\phi$ are arbitrary parameters. Thereby the same laser as in the
previous subsection can be used. To avoid the situation that the time evolution of the system is restricted to changes inside the decoherence-free
subspace, the atom should be moved out of the cavity.

If we neglect again spontaneous emission ($\Gamma=0)$, the laser
Hamiltonian which describes the time evolution of atom $i$ is given by
\begin{eqnarray}
H_{\rm laser \, I} &=& {\hbar \over 2} \, \Big( \,
\Omega^{(i)} \, |1\>_{ii} \<0| +  {\rm h.c.} \, \Big) ~.
\end{eqnarray}
Calculating the corresponding time evolution operator for a laser
pulse length $T$ leads to Eq.~(\ref{urot}) with
\begin{eqnarray}
\xi = {|\Omega^{(i)}| T \over 2} ~~&{\rm and}&~~
{\rm e}^{{\rm i} \phi} = {\Omega^{(i)} \over |\Omega^{(i)}|} ~.
\end{eqnarray}
To change the phase $\phi$, the phase of the Rabi frequency $\Omega^{(i)}$
has to be chosen very carefully, while $\xi$ can easily be varied
by varying the length $T$ of the pulse.

Again, for $\Gamma\neq 0$ a photon may be emitted spontaneously during the single qubit rotation which leads to a failure of the experiment and therefore to a further decrease of the success rate of the scheme to test Bell's inequality proposed here.

\subsection{State measurement on a single atom}

Whether an atom $i$ is in state $|0 \>_i$ or $|1 \>_i$  can be
measured with a very high precision following a proposal by Cook
\cite{Cook}. To do this, we make use
of a short strong laser pulse and an auxiliary level $2$. The probe
pulse couples one of the states, for instance the state $|0 \>_i$ to
state $|2\>_i$, and has the Rabi frequency $\Omega_2$. The spontaneous
decay rate of the auxiliary level is $\Gamma_2$. If the length
of the laser pulse, $T$, fulfills a minimum length,
\begin{eqnarray} \label{cond2}
T &\gg& {\rm max}\, \left\{ 1/\Gamma_2,\,\Gamma_2/\Omega_2^2
\right\}~,
\end{eqnarray}
the absence or occurrence of photons from the 0-2 transition
indicates whether the atom is found in state $|0 \>_i$ or $|1 \>_i$,
respectively. If the system is initially prepared in level 0 photons
are emitted until the end of the pulse. If the atom is in $|1
\rangle_i$ the laser has no effect on the atomic state and no photon
emissions will occur. For an arbitrary
state of the atom
\begin{eqnarray} \label{ex}
|\varphi \> &=& \alpha_0 \, |0 \rangle_i + \alpha_1 \, |1 \rangle_i
\end{eqnarray}
it has been shown by Beige and Hegerfeldt \cite{behe} that photons
are emitted with
probability $|\alpha_0|^2$ as predicted for an ideal measurement. The
proposition for this scheme to work is that the laser pulse is long
enough that an atom initially in state $|0 \rangle_i$ emits
definitively a photon which leads to condition (\ref{ex}). As
discussed in Ref.~\cite{behe} the precision of this measurement can
be very high, even if the efficiency of the detectors measuring the
photons from the 0-2 transition is very low. The population
difference between the two levels is given by
\begin{eqnarray}
\<\sigma_z^{(i)}\> = 1 - 2 \, |\alpha_0|^2
\end{eqnarray}
averaged over many runs.

\section{An improved scheme using four-level atoms}

To observe a violation of Bell's inequality the preparation of the maximally entangled state $|a\>$ should succeed with a probability above 71$\,\%$ in each run of the experiment. For this, as can be seen in Fig.~\ref{fig2}, the coupling constant $g$ has to be at least 100 times larger than the spontaneous decay rate $\Gamma$. This is difficult to achieve experimentally using optical frequencies, and has only been realised in micro cavities 
with circular Rydberg atoms coupled to a microwave cavity\cite{arno}. 

In the following we describe how this problem can be circumvented easily by making use of two 
additional atomic levels. They allow us to replace all transitions in the two-level system by 
Raman transitions. We show that the four-level atoms possess the same decoherence-free states 
as the two-level atoms described in Section IIC and again a weak laser pulse can be used to 
create entanglement between the atoms. We describe how to perform a single qubit rotation 
and how to measure the state of an atom.

\subsection{The preparation of the entangled state}

\noindent
\begin{minipage}{3.38truein}
\begin{center}
\begin{figure}[h]
\epsfig{file=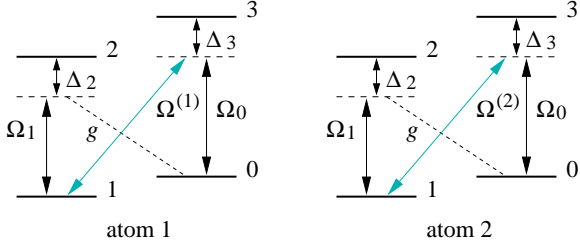,width=3truein} \\[0.2cm] 
\caption{The cavity mode couples with coupling strength $g$ and detuning 
$\Delta_2$ to the 0-2 transition of each atom. The 1-2 and the 0-3 transition are both 
driven by a laser field with Rabi frequency $\Omega_1$ or $\Omega_0$ and detuning 
$\Delta_2$ or $\Delta_3$, respectively. A weak laser field couples, in addition, to the
1-3 transition of atom $i$ with Rabi frequency $\Omega^{(i)}$ and detuning 
$\Delta_3$.}\label{lambda} 
\end{figure}
\end{center}
\end{minipage}
\vspace*{0.5cm}

We consider now two four-level atoms with a configuration as shown in Fig.~\ref{lambda}. 
The states $|0\>_i$ and $|1\>_i$ are the ground states of atom $i$ and couple to the excited 
states denoted by $|2\>_i$ and $|3\>_i$. Ground states and excited states could each be 
obtained from one degenerate level. To prepare the atoms in state (\ref{instate}) they 
have to be moved into a cavity as described in Section II. In the following, $\hbar \omega_i$ 
denotes the energy of level $i$. The frequency $\omega_{\rm cav}$ of the single cavity mode 
equals $\omega_{\rm cav} = \omega_2-\omega_0-\Delta_2$,
where $\Delta_2$ denotes a detuning. A laser field with the same detuning and frequency
$\omega_{21} = \omega_2-\omega_1-\Delta_2$ 
excites the 1-2 transition of each atom with Rabi frequency $\Omega_1$ and another laser 
drives the 0-3 transition of both atoms with Rabi frequency $\Omega_0$ and has the frequency
$\omega_{30} = \omega_3-\omega_0-\Delta_3$. In addition, at time $t=0$ a weak laser pulse 
with frequency $\omega_{31} = \omega_3-\omega_1-\Delta_3$ and with Rabi frequency $\Omega^{(i)}$ 
is applied to the 1-3 transition of atom $i$.

To describe the time evolution of the system  under the condition of no photon emission we 
use again the quantum jump approach \cite{HeWi1}. Here we chose the interaction picture with respect 
to the sum of the atomic Hamiltonian 
\begin{eqnarray} \label{h000}
H_0 &=& \sum_{i=1}^2 \sum_{j=0}^3 \hbar \omega_j \, |j\>_{ii}\<j| 
- \sum_{i=1}^2 \sum_{j=2}^3  \hbar \Delta_j \, |j\>_{ii}\<j| 
\end{eqnarray}
and the Hamiltonian describing the energy of the cavity mode and the free radiation fields 
forming the environment of the system. Then the conditional Hamiltonian becomes
\begin{eqnarray} \label{HcondL}
H_{\rm cond} &=&  {\rm i} \hbar \, g \, \sum_{i=1}^2 \Big( \, 
b \, |2\>_{ii} \<0| - {\rm h.c.} \, \Big) \nonumber \\ 
& & + {\hbar \over 2} \, \sum_{i=1}^2 \Big( \,
\Omega_1 \, |2\>_{ii}\<1| + \Omega_0 \, |3\>_{ii} \<0| + {\rm h.c.} \, \Big) \nonumber \\
& & + {\hbar \over 2} \, \sum_{i=1}^2 \Big( \,
\Omega^{(i)} \, |3\>_{ii} \<1| + {\rm h.c.} \, \Big) \nonumber \\
& &  - {\rm i} \hbar \sum_{i=1}^2 \sum_{j=2}^3 \left( \Gamma_j  + {\rm i}\Delta_j \right) 
|j\>_{ii} \<j| - {\rm i} \hbar \, \kappa \, b^\dagger b ~,
\end{eqnarray}
where we assumed again that the coupling constant $g$ is for both atoms the same and $\Gamma_j$
denotes the spontaneous decay rate of level $j$.

In the following we assume that the detunings $\Delta_2$ and $\Delta_3$ are much larger than 
all other system parameters, 
\begin{equation} \label{par}
|\Omega_0|~,~ |\Omega_1 | ~,~ |\Omega^{(i)}| ~,~ g ~,~ \Gamma_j ~ 
\ll \Delta_2 \sim \Delta_3 ~,
\end{equation}
and write the (unnormalized) state of the system under the condition of no photon emission 
as
\begin{eqnarray}
|\psi^0 (t) \> = \sum_{n=0}^\infty \sum_{j_1,j_2=0}^3 
c_{\rm n j_1j_2}(t) \, |nj_1j_2\> ~.
\end{eqnarray} 
Because we are only interested in the time evolution of the system on a time scale much 
longer than $1/\Delta_2$ and $1/\Delta_3$ level 2 and level 3 can be eliminated 
adiabatically by eliminating the fast varying coefficients. All coefficients with $j_1$ 
or $j_2$ equal to 2 or 3 adapt essentially immediately to the state of the other levels 
and we can set their derivatives in the Schr\"odinger equation corresponding to Eq.~(\ref{phit})
equal to zero. This allows us to determine the fast varying coefficients analytically. Substituting 
the result into the differential equations for the remaining slowly varying coefficients we
find that their time evolution under the condition of no photon emission is governed by the 
effective Hamiltonian $\tilde{H}_{\rm cond}$ with
\begin{eqnarray} \label{HcondLL}
\tilde{H}_{\rm cond} &=&  {\rm i} \hbar \, g_{\rm eff} \, \sum_{i=1}^2 \Big( \,
b \, |1\>_{ii} \<0| - {\rm h.c.} \, \Big) \nonumber \\ 
& & + {\hbar \over 2} \, \sum_{i=1}^2 \Big( \,
\Omega_{\rm eff}^{(i)} \, |1\>_{ii} \<0| + {\rm h.c.} \, \Big) 
- {\rm i} \hbar \, \kappa \, b^\dagger b \nonumber \\
& & - {\hbar \over 4} \sum_{i=1}^2 \Big( \,
{|\Omega_1|^2 \over \Delta_2} \, |1\rangle_{ii}\langle 1| 
+ {|\Omega_0|^2 \over \Delta_3} \, |0\rangle_{ii}\langle 0| \nonumber \\
& &
+ {|\Omega^{(i)}|^2 \over \Delta_3} \, |1\rangle_{ii}\langle 1|
+ {4 g^2 \over \Delta_2} \, b^\dagger b \, |0\rangle_{ii}\langle 0| \, \Big) ~.
\end{eqnarray}
Here all terms of second and higher order in $1/\Delta_2$ and $1/\Delta_3$ have been 
neglected. The effective atom-cavity coupling constant $g_{\rm eff}$ is given by
\begin{equation}
g_{\rm eff} \equiv - g \cdot\Omega_1^*/(2 \Delta_2) 
\end{equation}
and the effective Rabi frequencies $\Omega^{(i)}$ equal
\begin{equation} \label{eff}
\Omega^{(i)}_{\rm eff} \equiv - \Omega^{(i)} \cdot \Omega_0^*/(2 \Delta_3) ~.
\end{equation}
The level shifts in Eq.~(\ref{HcondLL}), which are proportional $1/\Delta_2$ and $1/\Delta_3$, 
can be neglected if they are for all states the same or if they are much smaller than the 
parameters governing the time evolution of the corresponding transition. We assume 
therefore in the following
\begin{equation} \label{par2}
|\Omega_0| = |\Omega_1| ~,~ \Delta_2 = \Delta_3 ~,~ g \ll |\Omega_1| 
~~{\rm and} ~~ |\Omega^{(i)}| \ll |\Omega_0| ~.
\end{equation}
For this parameter choice, the Hamiltonian $\tilde{H}_{\rm cond}$ resembles the conditional 
Hamiltonian (\ref{Hcond}) of the two-level atoms in Section II to a very good approximation.
Despite the values of $g$ and $\Omega^{(i)}$ are now replaced by $g_{\rm eff}$ and 
$\Omega_{\rm eff}^{(i)}$ and the spontaneous emission rate $\Gamma$ equals zero.

To prepare the atoms in the entangled state (\ref{instate}) therefore 
the same idea as in the previous section can be used. The decoherence-free states are exactly 
the same - the superpositions of the two states $|\rm 0g\>$ and $|\rm 0a\>$. 
In analogy to Section II, Eq.~(\ref{cond}), we assume now 
$|\Omega^{(i)}_{\rm eff}| \ll g_{\rm eff}$ and $\kappa \sim g_{\rm eff}$.
This leads in addition to Eq.~(\ref{par2}) to the condition
\begin{eqnarray} \label{par3} 
|\Omega^{(i)}| \ll g ~~ {\rm and} ~~ \kappa \sim g \Omega_1/\Delta_2 ~.
\end{eqnarray}
If condition (\ref{par2}) and (\ref{par3}) are fulfilled we expect that the weak 
laser pulse with the 
Rabi frequencies $\Omega^{(i)}$ does not move the system out of the decoherence-free subspace, 
if the system is initially in the ground state $|000\>$. Its effect can again be described by 
the effective Hamiltonian $H_{\rm eff}$ given in Eq.~(\ref{Heff2}). One only has to replace 
the Rabi frequencies $\Omega^{(i)}$ by $\Omega_{\rm eff}^{(i)}$.

\noindent
\begin{minipage}{3.38truein}
\begin{center}
\begin{figure}[h]
\epsfig{file=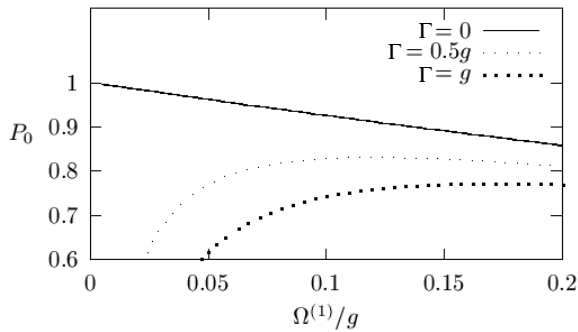,width=8cm}\\[0.3cm]
\caption{The probability for no photon emission during the
preparation of the maximally entangled state for different Rabi
frequencies $\Omega^{(1)}=-\Omega^{(2)}$ and $\Omega_0=\Omega_1=2 \, g$, 
$\Delta_2=\Delta_3 = 400 \, g$, $\kappa=0.0025 \, g$ and different spontaneous decay rates 
$\Gamma_2=\Gamma_3\equiv\Gamma$.}
\label{fig6}
\end{figure}
\end{center}
\end{minipage}
\vspace*{0.5cm}

Fig.~\ref{fig6} shows the probability for no photon emission for a laser pulse of the length 
$T = 2 \sqrt{2} \pi \Delta_3 /|\Omega_0(\Omega^{(1)}-\Omega^{(2)})|$
obtained from a numerical solution of Eq.~(\ref{phit})
with the Hamiltonian given in Eq.~(\ref{HcondL}). 
As expected from Eq.~(\ref{T}) and (\ref{eff}), the laser field prepares the atoms in this 
case in the 
maximally entangled state $|{\rm a}\>$. The fidelity is always higher than $99.5\,\%$ 
for the parameters chosen in Fig.~\ref{fig6}.
A comparison with Fig.~\ref{fig2} shows that the scheme behaves indeed to a very good approximation 
as the scheme in Section II but its success rate can now be very close to unity even if the 
spontaneous decay rates $\Gamma_j$ are of the same order of magnitude as the coupling constant 
$g$. 

\subsection{Realisation of a single qubit rotation}

To perform the single qubit rotation on atom $i$ the same three lasers as in the previous 
subsection with the Rabi frequencies $\Omega_0$, $\Omega_1$ and $\Omega^{(i)}$ can be used. 
As described in Section II, the atom has to be moved out of the cavity. Then all lasers 
are applied simultaneously. With respect to the interaction picture defined in 
Eq.~(\ref{h000}) the conditional Hamiltonian is now given by
\begin{eqnarray} \label{rot}
H_{\rm cond} 
&=& {\hbar \over 2} \, \Big( \,
\Omega_1 \, |2\>_{ii}\<1| + \Omega_0 \, |3\>_{ii}\<0| 
+ {\rm h.c.} \, \Big) \nonumber \\
& & + {\hbar \over 2} \, \Big( \, \Omega^{(i)} \, |3\>_{ii} \<1| 
+ {\rm h.c.} \, \Big) \nonumber \\
& & - {\rm i} \hbar \sum_{j=2}^3 (\Gamma_j + {\rm i} \Delta_j) \, |j\>_{ii}\<j| ~.
\end{eqnarray}
Eq.~(\ref{par}) allows us again to eliminate level 2 and level 3 adiabatically. 
Proceeding as in the previous subsection we find that the atom 
can effectively be described by the Hamiltonian
\begin{eqnarray} \label{shifts}
\tilde{H}_{\rm cond} &=& {\hbar \over 2} \, \Big( \,
\Omega^{(i)}_{\rm eff} \, |1\>_{ii}\<0| + {\rm h.c.} \, \Big) \nonumber \\
& & - {\hbar \over 4} \Big( \, {|\Omega_0|^2 \over \Delta_3} |0\>_{ii}\<0| 
+ {|\Omega_1|^2 \over \Delta_2} \, |1\rangle_{ii}\langle 1| \nonumber \\
& & + {|\Omega^{(i)}|^2 \over \Delta_3} \, |1\rangle_{ii} \langle 1| \, \Big) ~.
\end{eqnarray}
where Eq.~(\ref{eff}) has been used. 
This Hamiltonian does not depend on $\Gamma_2$ and $\Gamma_3$ and spontaneous emission 
by the atom can be neglected. If the parameters fulfill, as in the previous subsection, 
condition (\ref{par2}), then the last term in Eq.~(\ref{shifts}) is negligible
whilst the remaining level shifts are for all states the same and introduce an overall phase 
factor to the state of the atom. The time evolution operator corresponds therefore up 
to a total phase factor with the operator given in Eq.~(\ref{urot}) and equals
\begin{eqnarray} 
U(T,0) &=& \exp \left( {\rm i} \, {|\Omega_0|^2 T \over 4 \Delta_3} \right)
U^{(i)}_{\rm rot} (\xi, \phi)
\end{eqnarray}
with
\begin{eqnarray}
\xi = {|\Omega^{(i)} \Omega_0| T \over 4\Delta_3}  ~~&{\rm and}&~~
{\rm e}^{{\rm i} \phi} = -{\Omega^{(i)} \Omega_0^* \over |\Omega^{(i)} \Omega_0|} ~. 
\end{eqnarray}
We will see later that the additional phase factor does not affect the outcome of the 
Bell measurement described in the next section. We can therefore ignore this factor and 
use the Hamiltonian (\ref{rot}) to realise the single qubit rotation. 

\subsection{State measurement on a single atom}

To measure whether atom $i$ is in state $|0\>_i$ or $|1\>_i$, respectively, the same 
scheme as described in Section IIC can be used. 

\section{A test of the Bell inequality}

Given that the state (\ref{instate}) can be generated, the next
interesting question is whether such a state will violate one of
Bell's inequalities? For certain parameters it must but what physical
measurements are necessary to characterize this disagreement with
local realism?

\subsection{The Bell inequality}

The {\it spin} (or correlation function) Bell
inequality\cite{Bell65,CHSH69}
may be written formally as
\begin{eqnarray}\label{spinx}
B_{\rm S}=|E\left(\theta_{1},\theta_{2}\right)
&-&E\left(\theta_{1},\theta_{2}'\right) \nonumber \\
&+& E\left(\theta_{1}',\theta_{2}\right)
+ E\left(\theta_{1}',\theta_{2}'\right)|\leq 2~,
\end{eqnarray}
where the correlation function $E\left(\theta_{1},\theta_{2}\right)$
is given by
\begin{eqnarray}\label{correlation}
E\left(\theta_{1},\theta_{2}\right)&=&\< \sigma^{(1)}_{\theta_{1}}
\sigma^{(2)}_{\theta_{2}}\> ~.
\end{eqnarray}
Here $\theta_1$ and $\theta_2$ are real parameters.
In the following the operator $\sigma_a^{(i)}$ with $a=x$, $y$ or $z$
is the $a$ Pauli spin operators for the two-level system of atom $i$
and the operator $\sigma^{(i)}_{\theta_{i}}$ is defined as
\begin{equation} \label{yyy}
\sigma^{(i)}_{\theta_{i}}=\cos \theta_{i} \,
\sigma^{(i)}_x + \sin \theta_{i} \, \sigma^{(i)}_y~.
\end{equation}
We describe now how the inequality (\ref{spinx}) could be tested
experimentally.

\subsection{Description of the experimental test}

To test Bell's inequality the atoms have to be prepared first in a
state
for which a violation of Bell's inequality (\ref{spinx}) is expected.
This
can be done with the help of the scheme discussed in Section IIA by
preparing the atoms in state (\ref{instate}). The parameter $\alpha$
can
be varied by changing the length $T$ of the laser pulse.

For certain initial states and in certain cases (including here) the
correlation function depends only on the difference between the
angles
$\theta_{1}$ and $\theta_{2}$ and we have
\begin{eqnarray} \label{sss}
E\left(\theta_{1},\theta_{2}\right) &=&
E\left(\theta_{1}-\theta_{2},0 \right) ~.
\end{eqnarray}
This can be proven easily and holds because the state $|11\>$ is not
populated. Populating $|11\>$ by the preparation schemes proposed here
is not possible, because the time evolution of the system is restricted to
decoherence-free states \cite{arg}. As an example to test Bell's inequality we choose $\vartheta =
\theta_{1}-\theta_{2}=\theta_{2}-\theta_{1}'=
\theta_{1}'-\theta_{2}'$. This leads to
$\theta_{1}-\theta_{2}'=3\vartheta$.
Using Eq.~(\ref{sss}) the inequality (\ref{spinx}) simplifies for
this
parameter choice to
\begin{eqnarray} \label{bs}
B_{\rm S}=|3 E\left(\vartheta,0\right) - E\left(3
\vartheta,0\right)|\leq 2~.
\end{eqnarray}
A violation of this inequality corresponds to $|B_{\rm{S}}|>2$.

To find a way to measure the correlation functions
$E\left(\vartheta,0\right)$
we make use of the relation
\begin{eqnarray}
& & \hspace*{-0.7cm}
U_{\rm rot}^{(i)\, \dagger} (\xi,\phi) \, \sigma_z^{(i)} \,
U_{\rm rot}^{(i)} (\xi,\phi) \nonumber \\
~~ &=& \cos 2\xi \, \sigma_z^{(i)} - \sin 2\xi \, \left( \,
\cos \phi \, \sigma_y^{(i)} + \sin \phi \, \sigma_x^{(i)} \, \right)
~.
\end{eqnarray}
This allows us to rewrite $\sigma^{(i)}_{\theta_i}$ in terms of
$\sigma^{(i)}_z$. By choosing $\xi=\pi/4$ and by making use of some
trigonometric relations one obtains from Eq.~(\ref{yyy})
\begin{eqnarray} \label{xxxx}
\sigma^{(i)}_{\theta_i}  &=& U_{\rm rot}^{(i)\, \dagger}
\left({\pi \over 4},{3\pi \over 2}-\theta_i \right) \, \sigma_z^{(i)}
\,
U_{\rm rot}^{(i)} \left({\pi \over 4},{3\pi \over 2}-\theta_i \right)
~,
\end{eqnarray}
where $U_{\rm rot}^{(i)}$ is the single qubit rotation defined in
Eq.~(\ref{urot}). Using this, Eq.~(\ref{correlation}) and (\ref{sss})
one can show that
\begin{eqnarray} \label{xxx}
E(\vartheta,0) &=& \Bigg< U_{\rm rot}^{(1)\, \dagger}
\left({\pi \over 4},{3\pi \over 2}-\vartheta \right) \,
\sigma^{(1)}_{z} \,
U_{\rm rot}^{(1)} \left({\pi \over 4},{3\pi \over 2}-\vartheta \right)
\nonumber \\
& &  \times \,
U_{\rm rot}^{(2)\, \dagger} \left({\pi \over 4},{3\pi \over 2}
\right)
\, \sigma^{(2)}_{z} \, U_{\rm rot}^{(2)} \left({\pi \over 4},{3\pi
\over 2} \right) \Bigg>
\end{eqnarray}
This expectation value can be measured in the following way. First,
the single qubit rotation described in Section IIB has to be applied
on both
atoms with $\xi = \pi/4$ and $\phi =3\pi/2-\vartheta$ for atom 1 and
$\xi = \pi/4$
and $\phi =3\pi/2$ for atom 2. Afterwards the observables
$\sigma_z^{(1)}$
and $\sigma_z^{(2)}$ have to be measured. This can be done by
measuring
whether the atoms are in their ground state or not as described in
Section IIC or IIIC, respectively. In an analogous way $E(3\vartheta,0)$ can be determined experimentally.

It is important to point out that the correlation function represents
an ensemble average obtained by performing the measurements over many
runs, each time repreparing the initial state.

\subsection{Expected violation of Bill's inequality}

It is straightforward to show that the correlation function for the
initial state (\ref{instate}) is given by
\begin{eqnarray}
E\left(\vartheta,0\right)= - |\alpha|^{2} \cos \vartheta
\end{eqnarray}
and hence Eq.~(\ref{bs}) can assume a maximum of $|B_{\rm S}| = 2
\sqrt{2} \, |\alpha|^{2}$ where we have
chosen $\vartheta=\pi/4$. Therefore, a violation of the {\it spin}
Bell inequality is possible for $|\alpha|^{2} > 1/ \sqrt 2$.
The quantity $|\alpha|^2$ can be expressed in terms of the
fundamental system parameter $|\Omega^{(-)}| T$ only with the
help of
Eq.~(\ref{alpha}). In Fig (\ref{fig4}) we plot $|B_{\rm S}|$ versus
$|\Omega^{(-)}| T$ and $\vartheta$. 

\noindent
\begin{minipage}{3.38truein}
\begin{center}
\begin{figure}[h]
\center{\epsfig{file=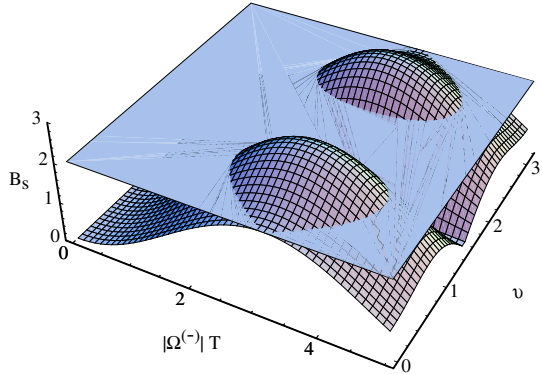,width=7.3cm}} \\[0.2cm]
\caption{Plot of $|B_{\rm S}|$ versus $|\Omega^{(-)}|T$ and
$\vartheta$.
A violation of the {\it spin} Bell's inequality occurs for $|B_{\rm
S}|>2$ and are displayed as {\it Islands} in the $|\Omega^{(-)}|T$
- $\vartheta$ plane. The angles have been chosen so as to maximise the violation utilising the maximally entangled state.}
\label{fig4}
\end{figure}
\end{center}
\end{minipage}
\vspace*{0.5cm}

A significant region of violation is observed with the maximum
of $|B_{\rm S}|=2\sqrt{2}$ occurring at $|\Omega^{(-)}|T = \pi$.
The state of the atoms at such a time is a maximally entangled
state.
This test on Bell's inequality should be feasible with current
technology.

\section{Discussion}

In this article we have made use of a recently proposed scheme \cite{ent}
to prepare in a controlled way with a very high success rate two atoms in an arbitrary superposition of a maximally entangled state and a product state. We show how the spin Bell inequality\cite{Bell65,CHSH69} can be characterised, tested and violated closing the detection loophole. To do so we use the highly efficient measurement proposal by Cook \cite{Cook} based on ``electron shelving''. The system discussed here has the appeal that the atoms are massive particles compared with photons and hence our proposal tests quantum mechanics in an all new macroscopic regime. In addition, while the photon experiments close the casualty loophole, the proposed atom experiment would close the detection efficiency loophole. Therefore, the experiment we discuss is complementary to the current photon experiments being performed.

{\it To summarise}, entanglement is a necessary quantum
resource used in quantum information. While entangled photons have
to date been the engine of much recent work, their `flying' nature
renders them inappropriate for the storage of information. We have
discussed a means in which trapped ions or atoms become entangled in a controlled way using dissipation, and the degree to which the resulting entanglement can be measured through Bell correlations.\\

\noindent {\em Acknowledgement:}
Part of this work was carried out at the Isaac Newton Institute for
Mathematical Sciences in Cambridge and we acknowledge their support
and hospitality as well as the support by the European Science
Foundation.
We acknowledge also the support of the UK Engineering and
Physical Sciences Research Council, the Australian Research Council,
the European Union and the Alexander von Humboldt Foundation. We
gratefully thank J. I. Cirac, W. Lange, G. J. Milburn, I. C. Percival, A. Rauschenbeutel, 
B. Varcoe, and K. W\'odkiewicz for valuable discussions and B. Tregenna for pointing out 
a mistake in Section III in an earlier verion of this paper.\\

\end{multicols}


\begin{references}
\small
\bibitem[\dagger]{AB} Electronic address: a.beige@ic.ac.uk
%
%
%
\bibitem{Bell65} 
J. S. Bell, Physics (N.Y.) {\bf 1}, 195 (1965).
%
\bibitem{Clauser and Horne 74} 
J. F. Clauser and M. A. Horne, Phys. Rev. D {\bf 10}, 526 (1974).
%
\bibitem{CHSH69} 
J. F. Clauser, M. A. Horne, A. Shimony and R. A. Holt, Phys. Rev. Lett. {\bf 23}, 880 (1969).
%
\bibitem{Braunstein and Caves 1988} 
S. L. Braunstein and C. M. Caves, Phys. Rev. Lett. {\bf 61}, 622 (1988).
%
\bibitem{Peres} 
A. Peres, Found. Phys. {\bf 29}, 589 (1999).
%
\bibitem{Clauser} 
J. F. Clauser and A. Shimony, Rep. Prog. Phys. {\bf 41}, 1981 (1978) and references therein.
%
\bibitem{Aspect} 
A. Aspect, P. Grangier, and G. Roger, Phys. Rev. Lett. {\bf 49}, 91 (1982);  
A. Aspect, J. Dalibard, and G. Roger, Phys. Rev. Lett. {\bf 49}, 1804 (1982).
%
\bibitem{KwiatI} 
P. G. Kwiat, K. Mattle, H. Weinfurter, A. Zeilinger, A. V. Sergienko, and Y. Shih, Phys. Rev. Lett. {\bf 75}, 4337 (1995).
%
\bibitem{Gisin} 
W. Tittel, J. Brendel, H. Zbinden, and N. Gisin,
Phys. Rev. Lett. {\bf 81}, 3563 (1998).
%
\bibitem{Kwait} 
P. G. Kwiat, E. Waks,  A. G. White, I. Appelbaum,
and P. H. Eberhard, Phys. Rev. A {\bf 60}, R773 (1999).
%
\bibitem{Weihs} 
G. Weihs, T. Jennewein, C. Simon, H. Weinfurter, and A. Zeilinger, Phys. Rev. Lett. {\bf 81}, 5039 (1998).
%
\bibitem{mandel}
A. Kuzmich, I. A. Walmsley, and L. Mandel, Phys. Rev. Lett. {\bf 85}, 1349 (2000).
%
\bibitem{ent} 
A. Beige, D. Braun, and P. L. Knight, New J. Phys. {\bf 2}, 22 (2000).
%
\bibitem{Cook} 
R. C. Cook, Physica Scr. T{\bf 21}, 49 (1998).
%
\bibitem{behe}
A. Beige and G. C. Hegerfeldt, J. Mod. Phys. {\bf 44}, 345 (1997).
%
\bibitem{Cirac} 
J. I. Cirac and P. Zoller, Phys. Rev. A {\bf 50}, R2799 (1994).
%
\bibitem{Phoenix} 
S. J. D. Phoenix and S. M. Barnett, J. Mod. Opt. {\bf 40}, 979 (1993).
%
\bibitem{Knight} 
I. K. Kudryavtsev and P. L. Knight, J. Mod. Opt. {\bf 40}, 1673 (1993).
%
\bibitem{Fry} 
E. S. Fry, T. Walther, and S. Li, Phys. Rev. A {\bf 52}, 4381 (1995). 
%
\bibitem{Gerry} 
C. C. Gerry, Phys. Rev. A {\bf 53}, 2857 (1996).
%
\bibitem{Hagley} 
E. Hagley, X. Maitre, G. Nogues, C. Wunderlich, M. Brune, J. M. Raimond, and S. Haroche, Phys. Rev. Lett. {\bf 79}, 1 (1997). 
%
\bibitem{Sackett}
C. A. Sackett, D. Kielpinski, B. E. King, C. Langer, V. Meyer, C. J. Myatt,
M. Rowe, Q. A. Turchette, W. M. Itano, D. J. Wineland, and I. C. Monroe,
Nature {\bf 404}, 256 (2000). 
%
\bibitem{Molmer}
K. M{\o}lmer and A. S{\o}rensen, Phys. Rev. Lett. {\bf 82}, 1835 (1999);
K. M{\o}lmer and A. S{\o}rensen, Phys. Rev. A {\bf 62}, 022311 (2000). 
%
\bibitem{strong} By a strong test of quantum we mean a test with no
auxiliary assumptions. A test in the spirit of Bell original
description.
%
\bibitem{HeWi1} 
G. C. Hegerfeldt and T. S. Wilser, in {\em Classical and Quantum Systems}, Proceedings of the Second International Wigner Symposium, July 1991, edited by H. D. Doebner, W. Scherer, and F. Schroeck (World Scientific, Singapore, 1992), p. 104; G. C. Hegerfeldt and D. G. Sondermann,
Quantum Semiclass. Opt. {\bf 8}, 121 (1996).
%
\bibitem{HeWi2}
J. Dalibard, Y. Castin, and K. M{\o}lmer, Phys. Rev. Lett. {\bf 68}, 580 (1992).
%
\bibitem{HeWi3}
H. Carmichael, {\em An Open Systems Approach to Quantum Optics}, Lecture Notes in Physics, Vol. {\bf 18} (Springer, Berlin, 1993).
%
\bibitem{HeWi4}
For a recent review see M. B. Plenio and P. L. Knight,
Rev. Mod. Phys. {\bf 70}, 101 (1998) and references therein.
%
\bibitem{alt}
M. B. Plenio, S. F. Huelga, A. Beige, and P. L. Knight, Phys. Rev. A
{\bf 59}, 2468 (1999).
%
\bibitem{trapped1} 
P. M. Radmore and P. L. Knight, J. Phys. B {\bf 15}, 561 (1982).
%
\bibitem{trapped2}
G. M. Meyer and G. Yeoman, Phys. Rev. Lett. {\bf 79}, 1650 (1997).
%
\bibitem{trapped3}
G. J. Yang, O. Zobay, and P. Meystre, Phys. Rev. A {\bf 59}, 4012
(1999).
%
\bibitem{DFS1} 
G. M. Palma, K. A. Suominen, and A. K. Ekert, Proc.
Roy. Soc. London Ser. A {\bf 452}, 567 (1996).
%
\bibitem{DFS2} 
P. Zanardi and M. Rasetti, Phys. Rev. Lett. {\bf 79}, 3306 (1997).
%
\bibitem{DFS3} 
D. A. Lidar, I. L. Chuang, and K. B. Whaley, Phys. Rev. Lett. {\bf 81}, 2594
(1998).
%
\bibitem{misra} 
B. Misra and E. C. G. Sudarshan, J. Math. Phys. {\bf 18}, 756 (1977).
%
\bibitem{Itano} W. M. Itano, D. J. Heinzen, J. J. Bollinger, and
D. J. Wineland, Phys. Rev. A {\bf 41}, 2295 (1990).
%
\bibitem{zeno} A. Beige and G. C. Hegerfeldt, Found. Phys. {\bf 263},
1671 (1997).
%
\bibitem{messung}
A. Beige, S. Bose, D. Braun, S. F. Huelga, P. L. Knight,
M. B. Plenio, and V. Vedral, J. Mod. Opt. {\bf 47}, 2583 (2000).
%
\bibitem{rab}
The scheme works as well for
arbitrary choices of the Rabi frequencies as long as one has
$\Omega^{(1)} \neq \Omega^{(2)}$. It is also possible to apply a laser pulse
on one atom only.
%
\bibitem{arno}
M. Brune, F. Schmidt-Kaler, A. Maali, J. Dreyer, E. Hagley, J. M. Raimond, and S. Haroche, Phys. Rev. Lett. {\bf 76}, 1800 (1996).  
%
\bibitem{pop} K. Bergmann, H. Heuer, and B. W. Shore, Rev. Mod. Phys. {\bf 70}, 1003 (1998).
%
\bibitem{Vit} N. V. Vitanov and S. Stenholm, Phys. Rev. A {\bf 55}, 648 (1997).
%
\bibitem{arg}
If the state $|11\>$ would be populated its population vanishes
immediately during the next time interval $\Delta T$ by leakage of a
photon out of the cavity or due to the no-photon time evolution with
$H_{\rm cond}$ given in Eq.~(\ref{Hcond}).


\end{references}
\end{document}